# Competing Hydrogenation Pathways to Metastable CaH$_6$ Revealed by Machine-Learning-Potential Molecular Dynamics

Ryuhei Sato[1,*], Peter I. C. Cooke[2,†], Maélie Caussé[2,3], Hung Ba Tran[3], Seong Hoon Jang[3], Di Zhang[3], Hao Li[3], Shin-ichi Orimo[3,4], Yasushi Shibuta[1], Chris J. Pickard[2,3]

[1]Department of Materials Engineering, The University of Tokyo, Bunkyo-ku, Tokyo, Japan.
[2]Department of Materials Science and Metallurgy, University of Cambridge, Cambridge, United Kingdom
[3]Advanced Institute for Materials Research (WPI-AIMR), Tohoku University, Sendai, Miyagi, Japan
[4]Institute for Materials Research, Tohoku University, Sendai, Miyagi, Japan

*Ryuhei Sato: r-sato@g.ecc.u-tokyo.ac.jp

†Peter I. C. Cooke: pc729@cam.ac.uk

**ABSTRACT**.
The synthesis of the high-$T_c$ superhydride $CaH_6$ has stimulated significant interest in understanding synthesis pathways for metastable hydrides. However, the microscopic mechanisms governing such hydrogenation reactions remain poorly understood. Here, we show that machine-learning potential molecular dynamics (MLP-MD) simulations can reproduce and distinguish competing reaction pathways leading to metastable and stable hydrides. By simulating hydrogenation reactions at $CaH_2/H_2$ and $CaH_4/H_2$ interfaces, we identify two distinct pathways that produce clathrate-type $CaH_6$ and A15-type $CaH_{5.75}$, respectively. $CaH_{5.75}$ lies on the convex hull but requires extensive Ca sublattice rearrangement and therefore forms only at elevated temperatures. In contrast, $CaH_6$ becomes kinetically accessible when $CaH_2$ is used as the precursor. The crystallographic compatibility between the Ca sublattice of $CaH_2$ and the bcc framework of $CaH_6$ enables a martensitic-like topotactic transformation that bypasses the reconstructive pathway leading to $CaH_{5.75}$. These results reveal how precursor structure and thermodynamic stability compete to determine superhydride formation pathways and demonstrate that machine-learning molecular dynamics can directly capture the kinetic selection of metastable phases in reactive materials systems.

## I. INTRODUCTION.

Since the discovery of superconductivity in mercury by K. Onnes in 1911[1], achieving superconductivity at room temperature has remained a central challenge in condensed-matter physics. While superconductors underpin technologies such as quantum computing and magnetic resonance imaging, all known materials operate far below room temperature, limiting their widespread practical use.

Among known superconducting materials, superhydrides[2,3] exhibit the highest superconducting transition temperatures ($T_c$) and are regarded as promising candidates for realizing room-temperature superconductivity. For example, $CaH_6$[4,5,6] has been reported to exhibit a $T_c$ of approximately 210 K at 160 GPa. Although such high $T_c$ values have been achieved in binary hydride systems[7,8,9,10,11,12], room-temperature superconductivity has not yet been realized, motivating the exploration of multicomponent hydride systems[13,14,15,16]. Indeed, while independent experimental verification is still required, reports of room-temperature superconductivity have emerged in ultrahigh-pressure ternary systems such as La–Sc–H, with $T_c$ of 271–298 K at pressures of 195–266 GPa[17]. However, such extreme pressures pose significant challenges for practical applications, prompting growing interest in multicomponent polyhydrides that may exhibit high $T_c$ at more moderate pressures[18,19,20].

Advances in crystal-structure prediction, driven by the development of machine-learning workflows[21,22] and universal machine-learning potentials[23], have dramatically accelerated the discovery of candidate superconducting hydrides. Nevertheless, substantial challenges remain in assessing whether predicted structures are experimentally synthesizable and, more importantly, in identifying viable synthesis pathways. For instance, theoretical studies on metastable $Mg_2IrH_6$ proposed a synthesis route via $Mg_2IrH_7$ based on thermodynamic stability and structure prediction[18]. Subsequent experiments, however, reported the formation of a disordered $Mg_2IrH_5$ phase instead[19], highlighting the difficulty of predicting synthesis pathways based solely on thermodynamic considerations. In our previous work, we successfully reproduced hydrogenation reactions leading to polyhydrides such as $CaH_4$ using machine-learning-potential molecular dynamics (MLP-MD) simulations[24]. In principle, such simulations enable the direct observation of reaction pathways and kinetic competition between multiple products under extreme conditions. However, there are currently no clear guidelines for the targeted synthesis of metastable phases (i.e., thermodynamically unstable phases with respect to the 0 K convex hull). Although stable polyhydrides are expected to be readily obtained at high pressures due to reduced activation energies, as suggested in our previous study, metastable phases remain difficult to access in a controlled manner. Such metastable phases have attracted increasing attention because some of them are associated with high superconducting transition temperatures, as in $Mg_2IrH_6$ and $CaH_6$.

In this study, we focus on metastable $CaH_6$, a superhydride whose synthesis required nearly a decade[5] after its theoretical prediction. Using MLP-MD simulations, we investigate superhydride formation reactions at 150 GPa to explore feasible synthesis routes for metastable superhydrides. Based on the elucidated reaction mechanisms, we discuss the physicochemical factors governing the formation of metastable superhydrides and propose insights into practical synthesis strategies.

## II. METHODS

### A. Training data collection by DFT calculation

Approximately 0.15 million DFT trajectories were employed as the training dataset as summarized in Table S1. The dataset was curated to ensure transferability across the wide range of atomic environments that occur along the synthesis pathway.

*Ryuhei Sato: r-sato@g.ecc.u-tokyo.ac.jp

†Peter I. C. Cooke: pc729@cam.ac.uk

Firstly, we collected the DFT-MD data for Ca, $CaH_2$, $CaH_4$, $CaH_6$, $CaH_{24}$, and $H_2$ so that structures with various stoichiometry during the hydrogenation reactions are explicitly included, meaning that all dynamics occurs in an interpolated region of the constructed MLP. DFT-MD data at high temperatures (3,000-10,000 K) ensure the sampling of unstable structures, such as the liquid phase, under high pressure. This also enables the MLP to accurately reproduce hydrogenation reactions that occur via the liquid phase (i.e. surface melting-driven hydrogenation reaction[24]).

Additionally, DFT-MD simulations of heterogeneous cells for $CaH_2(100)/H_2$, $CaH_4(001)/H_2$, and $CaH_6(100)/H_2$ under high pressure were included to improve the accuracy of atomic behavior near the interface. DFT-MD data of metallic hydrogen under 600 GPa was also included, to account for possible molecular-atomic decomposition of $H_2$ under HT and HP conditions, as is known to occur at 2000 K and 100 GPa[25].

To efficiently learn short-range interatomic repulsion [26], additional datasets were included consisting of geometry optimizations of systems compressed by 20% from the most stable structures. Finally, to enable the MLP to capture pressure-dependent behavior, DFT calculations of optimized structures of Ca, $CaH_2$, $CaH_4$, $CaH_6$, $CaH_{24}$, and $H_2$ at pressures ranging from 0 to 300 GPa were also incorporated into the training dataset in addition to those listed in Table S1[30].

All these DFT calculations were performed using the Vienna Ab initio Simulation Package (VASP)[27] with the Perdew–Burke–Ernzerhof (PBE) exchange–correlation functional [28]. Standard PBE PAW potentials [29] were used to represent the core states, whereas the valence states (Ca $3p^64s^2$, and H: $1s^1$) were treated explicitly by the plane-wave basis-set with an energy cutoff of 400 eV. 2×2×2 k-point meshes were employed for the Brillouin zone sampling in most of the calculations listed in Table S1[30].

### B. Machine learning potential construction

The MLP was trained using the Deep Potential-Smooth Edition scheme as implemented in the DeePMD-kit package [31]. We chose an embedding network with three hidden layers and (25, 50, 100) nodes per layer. The size of the embedding matrix was set to 16. Three hidden layers with (240, 240, 240) nodes per layer were used in the fitting network. The cut-off radius was set to 6.0 Å and the descriptors decay smoothly from 0.35 to 6.0 Å.

The mean absolute error (MAE) of atomic forces between DFT and MLP calculations was around 0.2 eV/Å for training and testing data below 3000 K and 150 GPa, which is comparable to previous studies of MLP for finite temperature and high pressure simulations[24,32]. The constructed MLP also reproduces atomic forces during MLP-MD simulations for $CaH_{5.75}$ and $CaH_{12}$ with comparable accuracy, despite no DFT data for either of these structures being present in the training dataset, demonstrating the robustness and transferability of the constructed MLP.

### C. Molecular dynamics simulation

The Large-scale Atomic/Molecular Massively Parallel Simulator (LAMMPS[33]) was used for MD simulations. A Nose–Hoover thermostat [34] and a Parrinello-Rahman barostat [35] were used for the NPT-MD simulations. The timestep was set to 1 fs and the mass of deuterium ($D$) was used instead of that of hydrogen.

In this study, we used two different $CaH_x/H_2$ interfaces for hydrogenation reactions (see FIG. S1 in supporting information[30] for the snapshots). One is $CaH_4(101)/H_2$ interface containing about 8000 atoms (Ca, 1152 atoms; H, 7008 atoms). The other is $CaH_2(100)/H_2$ with 5000 atoms (Ca, 576 atoms; H, 4596 atoms). The initial cell parameters are 16.9 × 16.8 × 95.6 Å$^3$ and 11.3 × 16.1 × 88.9 Å$^3$, which are sufficiently large compared with the MLP cutoff radius (6 Å). For these interface systems, we initialized MLP-MD simulations at 300 K and heated for about 50 ps until the system reaches the specific temperature under fixed pressure (150 GPa), as shown in Fig. S2[30]. Ovito[36] was used for visualization of the MLP-MD simulation results.

### III. RESULTS
### A. A15-type $CaH_{5.75}$ superhydride synthesis

Figure 1 shows the snapshots of the atomic configurations during MLP-MD simulations of $CaH_4/H_2$ interface at 1500 K and 150 GPa. After approximately 40 ps, Ca atoms start to dissolve into the $H_2$ high-pressure phase, forming a solid solution (referred to as $H_2(Ca)$). Notably, the local structure around Ca atoms in this phase resembles a $CaH_{12}$ phase, a phase composed of Ca coordinated by $H_2$ dimers, which has been reported to be stable above 100 GPa[4].

The radial distribution functions (RDFs) of the Ca-dissolved $H_2$ phase are shown in Fig. S3. The observed

*Ryuhei Sato: r-sato@g.ecc.u-tokyo.ac.jp

†Peter I. C. Cooke: pc729@cam.ac.uk

RDF peaks coincide with those of $CaH_{5.75}$ and $CaH_{12}$. This indicates that the $H_2$ phase accommodates Ca atoms and forms a liquid-like Ca–H environment characterized by local structural motifs similar to these phases.

The recurrence of similar structural motifs in both a molten phase and $CaH_x$ phases (i.e. $CaH_{5.75}$ and $CaH_{12}$) that were not included in the training data is noteworthy for two reasons. Firstly, as we have explicitly validated the structural description of the $CaH_{5.75}$ and $CaH_{12}$ phases using DFT-MD as an independent reference for our MLP, it is likely that the appearance of these motifs in the molten phase is also valid and physically sensible. Secondly, the similarity in structural motifs hints at possible kinetically favored pathways for crystallization from the melt.

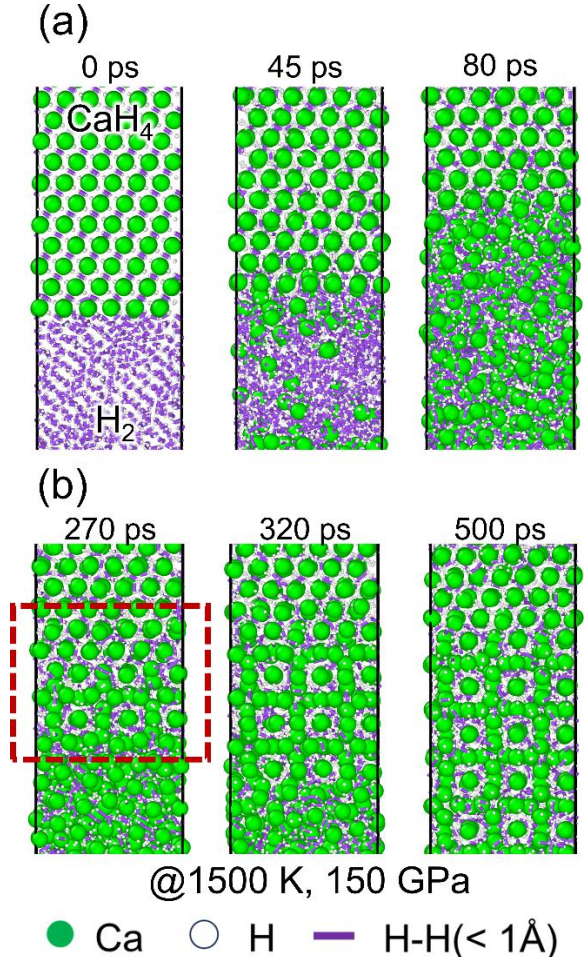

**FIG 1.** Snapshots of the atomic configurations of $CaH_4/H_2$ interface during MLP-MD simulations at 1500 K and 150 GPa. (a) the initial Ca dissolution into $H_2$ layers (b) $CaH_{5.75}$ formation near the interface after Ca supersaturation. The corresponding snapshots of the whole simulation cell are shown in Fig. S4[30].

After the $H_2$ phase becomes saturated with Ca and sufficient time has elapsed, a new bulk $CaH_x$ phase nucleates at the interface, as highlighted by the red dashed box at $t = 270$ ps in Fig. 1. This $CaH_x$ region subsequently grows with time and becomes the dominant phase within the simulation cell ($t = 500$ ps). Since the system forms a liquid-like phase ($H_2(Ca)$) as an intermediate state before forming a new superhydride, this hydrogenation reaction is successfully explained by surface-melting-driven hydrogen absorption -- our previous thermodynamic model for polyhydrides[24].

Figure 2a shows a snapshot after 1 ns of MLP-MD at 150 GPa and 1500 K, along with the resulting atomic structure, and the corresponding XRD diffraction pattern compared to that calculated for A15-type $CaH_{5.75}$. After 1 ns of simulation, all Ca atoms in the MLP-MD calculation shown in Fig. 1 aggregate to form a bulk phase with an A15-type framework. As confirmed by the XRD diffraction pattern, this structure is consistent with that of $CaH_{5.75}$. Note that some XRD peaks appear below ~10° (Cu $K\alpha$) in the MLP-MD simulation result. These low-angle peaks originate from hydrogen-enriched grain-boundary regions in the simulation cell. This is a consequence of a fixed overall H:Ca ratio of 6:1 in the MLP-MD cells. Nevertheless, the RDF obtained from $CaH_{5.75}/H_2$ interface agrees well with that of $CaH_{5.75}$ (Fig. 2(c)), indicating that the hydrogen content within the bulk phase is essentially equivalent to that of $CaH_{5.75}$.

This result appears to contradict previous experimental work, which suggested that the $CaH_6$ phase was synthesized at these conditions. However, reanalysis of the obtained XRD patterns suggests that $CaH_{5.75}$ was also present in the final product. Figure 3 compares the experimental XRD diffraction pattern of $CaH_x$ at 173 GPa reported by L. Ma et al.[5] with that calculated for $CaH_{5.75}$ after structural optimization at 160 GPa using DFT. Diffraction peaks observed near $\theta = 13°$ and $16°$ that were previously assigned to an unknown phase are found to coincide well with the calculated peak positions of $CaH_{5.75}$.

*Ryuhei Sato: r-sato@g.ecc.u-tokyo.ac.jp

†Peter I. C. Cooke: pc729@cam.ac.uk

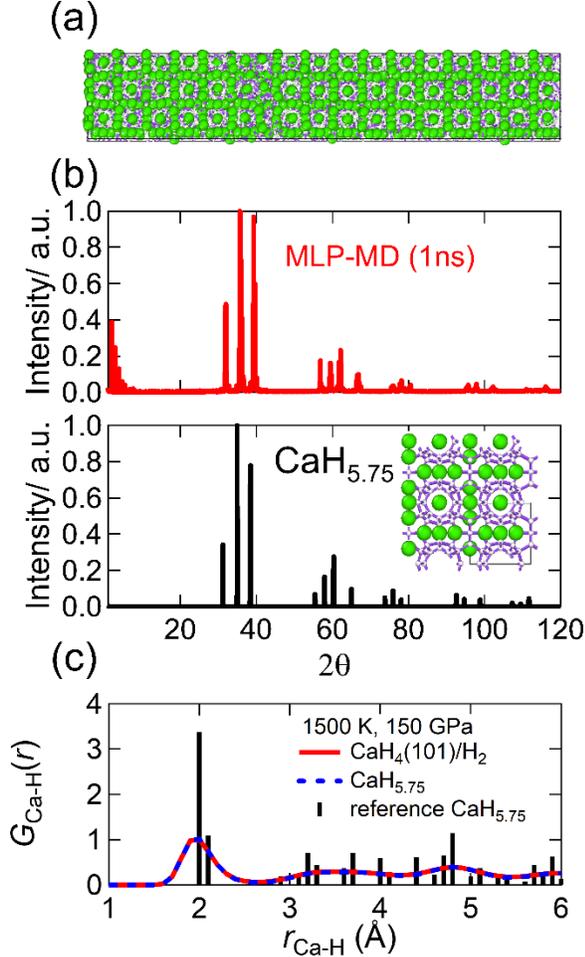

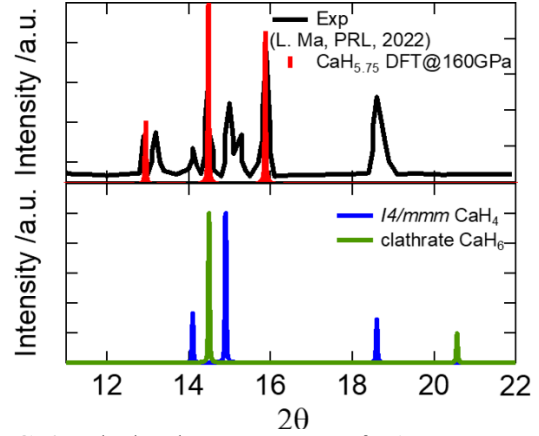

**FIG. 2.** (a) Snapshots of atomic configurations of the whole simulation cell for the CaH$_4$/H$_2$ interface obtained from 1 ns MLP-MD simulations at 1500 K and 150 GPa. (b) Calculated XRD patterns of (top) the structure obtained from the MLP-MD simulation and (bottom) the reference A15-type CaH$_{5.75}$ structure (a = 5.53 Å) optimized by DFT at 150 GPa. (c) Time-averaged radial distribution functions (RDFs) of H atoms around Ca atoms obtained from the MLP-MD simulation of the CaH$_4$/H$_2$ interface after CaH$_{5.75}$ formation (0.9–1.0 ns). The black solid line represents the RDF of the reference CaH$_{5.75}$ structure, while the blue dashed line corresponds to the MLP-MD simulation of bulk CaH$_{5.75}$ at 1500 K and 150 GPa.

In addition, the formation energy of CaH$_6$ from CaH$_{5.75}$ (i.e. CaH$_{5.75}$ + H$_2$ → CaH$_6$) is calculated to be +0.29 eV per formula unit. Therefore, from a thermodynamic point of view, CaH$_{5.75}$ is more stable than CaH$_6$ under these conditions. Moreover, A15-type superhydrides similar to CaH$_{5.75}$ have been reported to be stable in other systems[37,38,39] such as Ba

*Ryuhei Sato: r-sato@g.ecc.u-tokyo.ac.jp

†Peter I. C. Cooke: pc729@cam.ac.uk

and La, suggesting that the appearance of an A15 structure in the Ca system is not unexpected.

**FIG. 3.** calculated XRD patterns of A15-type CaH$_{5.75}$ structure (a = 5.50 Å) optimized by DFT at 160 GPa. The black line is the experimental XRD pattern taken from the supporting information (Fig. S2) of previous report by L. Ma et. al[5], while the blue and green lines are XRD patterns for *I4/mmm* CaH$_4$ (a = 2.71 Å, c = 5.05 Å) and clathrate CaH$_6$ (a = 3.47 Å), respectively.

### B. CaH$_6$ superhydride synthesis

Despite having a positive formation enthalpy with respect to CaH$_{5.75}$ and H$_2$, in this section we demonstrate that CaH$_6$ can be readily synthesized by selecting different precursors. Figure 4 shows snapshots from the MLP-MD simulations of the CaH$_2$(100)/H$_2$ interface at 1000 K and 150 GPa. Under these conditions, CaH$_x$ phases other than CaH$_{5.75}$ were formed after 1 ns, as illustrated in Fig. 4(a). Figure 4(b) shows a magnified view of this new CaH$_x$ bulk. As shown in the snapshot at 1000 ps, a hydrogen network similar to that of the clathrate CaH$_6$ was formed in this new CaH$_x$ bulk.

During this hydrogenation process, Ca atoms underwent small in-plane rearrangements, accompanied by an hcp-like to bcc-like transformation of the Ca sublattice. The Ca(100) plane of the *Pnma* CaH$_2$ structure corresponds to the (11-20) plane of *hcp* Ca, when considering only the Ca sublattice. The observed Ca rearrangement satisfies the orientation relationship (11-20)$_{hcp}$//(101)$_{bcc}$. This indicates that the hydrogenation reaction proceeds via a diffusionless transformation analogous to a martensitic-like phase transition. The RDF of the obtained CaH$_x$ bulk was comparable to that of the CaH$_6$ clathrate during MLP-MD simulation at 1000 K and 150 GPa. This demonstrates that both metastable CaH$_6$ and stable CaH$_{5.75}$ phases can be synthesized via high-pressure hydrogenation through appropriately designing the precursor and temperature conditions.

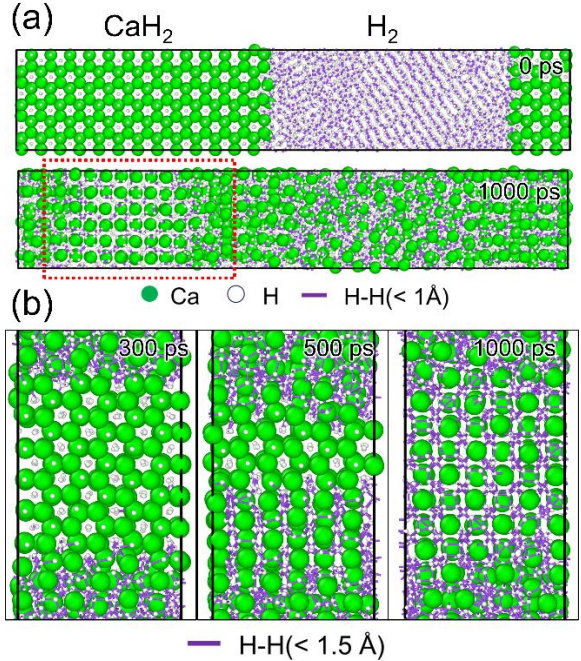

FIG. 4. (a) Snapshots of the atomic configurations during MLP-MD simulation of $CaH_2(100)/H_2$ interface at 1000 K and 150 GPa. (b) Magnified figures for the obtained new $CaH_x$ inside the slab. Longer H-H bonds (1.5 Å) were used for the clarity of hydrogen network in the clathrate $CaH_6$.

We note that, even in the MLP-MD simulations of the $CaH_2(100)/H_2$ interface at 1000 K, structures similar to $CaH_{5.75}$ were formed near the interface. A possible reason is that Ca atoms dissolve into the $H_2$ region, thereby enhancing Ca mobility near the interface.

This observation suggests that the obtained $CaH_x$ phase can be strongly affected by the overall composition of the system and the method by which hydrogen is supplied to the precursor. For example, under hydrogen-rich conditions such as those assumed in our MLP-MD simulations, $CaH_{5.75}$ may preferentially form. In contrast, when hydrogenation proceeds via hydrogen diffusion through polycrystalline grain boundaries of Ca metal or $CaH_x$, the space available for Ca diffusion and rearrangement is likely to be more restricted. Under such conditions, Ca rearrangement may be suppressed, and in combination with hydrogen deficiency (when compared to the $CaH_x/H_2$ interface), reaction pathways leading to the formation of $CaH_6$ may be relatively favored.

*Ryuhei Sato: r-sato@g.ecc.u-tokyo.ac.jp

†Peter I. C. Cooke: pc729@cam.ac.uk

These results indicate that the superhydride phase obtained can be sensitive to experimental conditions, with different reaction pathways being favored depending on the hydrogen supply and local structural environment. Accordingly, careful consideration is required when interpreting $CaH_x$ phases observed in the vicinity of interfaces.

## IV. DISCUSSION
### A. Stability and Accessibility of $CaH_6$ and A15-type $CaH_{5.75}$ phases

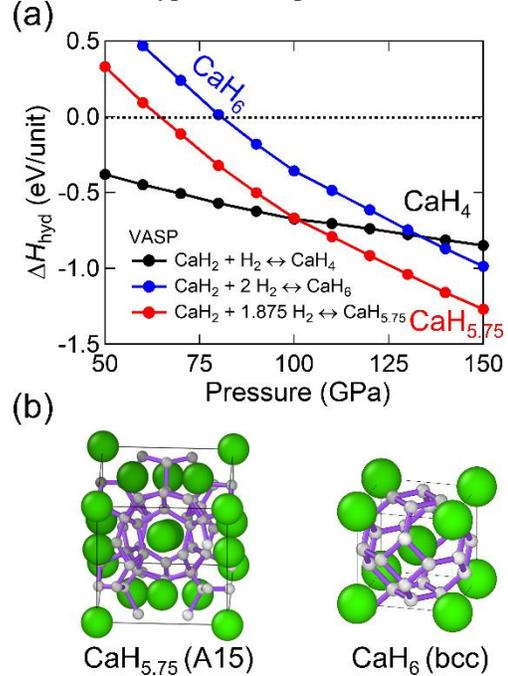

FIG. 5. (a) Hydrogenation enthalpy of $CaH_4$, $CaH_{5.75}$, and $CaH_6$ obtained from DFT calculation as a function of pressure. The same trends were obtained in MLP as summarized in FIG. S5[30]. (b) Structure of A15-$CaH_{5.75}$ and of clathrate $CaH_6$.

The origin of these two competing hydrogenation pathways can be attributed to a combination of the thermodynamic stability of $CaH_x$ phases and the structure of the Ca matrix in the precursor. Figure 5(a) shows the pressure dependence of the formation enthalpy of $CaH_x$ phases. Here, the formation enthalpy $\Delta H_{\text{hyd}}$ is defined with respect to $CaH_2$ and $H_2$; a negative value of $\Delta H_{\text{hyd}}$ indicates that the corresponding phase is more stable than $CaH_2$. At each pressure, the phase with the lowest $\Delta H_{\text{hyd}}$ is therefore thermodynamically favored. Although the results shown in Fig. 5(a) are obtained from DFT calculations, MLP employed in this study reproduces the same qualitative trends with good accuracy, as demonstrated in Fig. S5[30].

CaH$_{5.75}$ becomes thermodynamically more stable than CaH$_4$ above 100 GPa (Fig. 5(a)). In contrast, CaH$_6$ exhibits a higher formation enthalpy than other CaH$_x$ phases over the entire pressure range considered and is therefore a metastable phase. From a purely thermodynamic equilibrium perspective, the synthesis of CaH$_6$ is thus intrinsically unfavorable.

However, CaH$_{5.75}$ formation is kinetically suppressed. As illustrated in Fig. 5(b), the A15-type CaH$_{5.75}$ phase possesses a complex Ca atomic arrangement that differs significantly from typical bcc, fcc, or hcp crystal structures. Consequently, the formation of CaH$_{5.75}$ from CaH$_4$ with a bcc-type Ca matrix requires substantial rearrangement of Ca atoms, which requires high temperatures to enable sufficient Ca diffusion.

By contrast, the Ca sublattice of CaH$_6$ is bcc and exhibits high structural similarity to those of CaH$_2$ (hcp) and CaH$_4$ (bcc). As a result, CaH$_6$ can form without large Ca rearrangement. Nevertheless, the formation-enthalpy difference between CaH$_4$ and CaH$_6$ is very small (Fig. 5(a)), which limits the thermodynamic driving force to overcome the activation barrier for the hydrogenation reaction. Consistently, our MLP-MD simulations show that CaH$_6$ does not form at CaH$_4$(101)/H$_2$ interfaces under relatively low-temperature conditions around 1000 K. In the experimental study[5] (see Fig. 3 for the corresponding XRD pattern), CaH$_4$ and CaH$_{5.75}$ were reported to coexist in the probed regions. While the spatial distribution of these phases within the X-ray–irradiated region cannot be resolved, the XRD pattern measured in those regions does not show a clear reflection near ~20°, where CaH$_6$ is expected to exhibit a characteristic peak. This suggests that the contribution of CaH$_6$ in the probed region may be limited. Such an interpretation is compatible with our MLP-MD results, which indicate that CaH$_6$ formation from CaH$_4$ is kinetically hindered at moderate temperatures.

In the case of a CaH$_2$ precursor, the formation-enthalpy difference of approximately 1 eV/unit exists at 150 GPa, providing a sufficient thermodynamic driving force for hydrogenation. Owing to the combination of this large driving force and the high structural similarity of the Ca matrix, CaH$_6$ formation is observed at CaH$_2$(100)/H$_2$ interfaces even under low-temperature conditions where the formation of CaH$_{5.75}$ is kinetically suppressed.

*Ryuhei Sato: r-sato@g.ecc.u-tokyo.ac.jp

†Peter I. C. Cooke: pc729@cam.ac.uk

Experimentally, both CaH$_6$ and CaH$_{5.75}$ have been reported to form under nominally similar conditions around 2000 K and 173–190 GPa, as evidenced by Fig. 1 and Fig. S2 in Ma *et al.*, Phys. Rev. Lett. (2022)[5]. This suggests that, even experimentally, the observed phase formation is influenced not only by thermodynamic equilibrium but also by kinetic constraints.

In previous experiments, CaH$_x$ phases were synthesized under laser heating at temperatures of approximately 2000 K. In our MLP-MD simulations, as summarized in Supporting Information Section S1 and Fig. S7, CaH$_6$ loses its crystalline order and enters a liquid-like state at around 1750 K within 1 ns simulation. At lower temperatures, such as 1500 K, CaH$_6$ does not maintain its bcc structure but instead undergoes a structural transformation toward a different phase, likely related to CaH$_{5.75}$, as illustrated by the snapshots in Fig. S7. CaH$_{5.75}$ also exhibits liquid-like behavior upon heating to around 2000 K. The higher melting temperature of CaH$_{5.75}$ compared with CaH$_6$ in these MLP-MD simulations is consistent with the greater thermodynamic stability of CaH$_{5.75}$ relative to CaH$_6$.

These observations suggest that laser heating is likely to generate a transient liquid-like Ca–H environment in the vicinity of the heated region, rather than stabilizing well-defined crystalline CaH$_x$ phases at high temperature. It is also noteworthy that, in the high-temperature limit (~2000 K), the stability of CaH$_6$ and CaH$_{5.75}$ becomes effectively equivalent, since both systems evolve into a liquid-like state with an H:Ca ratio close to 6:1.

Consequently, it may be necessary to consider hydrogenation reactions occurring at heterogeneous interfaces under conditions with significant temperature gradients, as well as bulk crystallization processes during rapid thermal quenching after laser irradiation. Detailed information of the temperature history of the sample, particularly in regions where XRD diffraction patterns are measured, would facilitate a much more definitive discussion.

A recent study has also discussed the stability of CaH$_{5.75}$ and CaH$_6$[40]. CaH$_{5.75}$ lies on the convex hull, whereas CaH$_6$ is metastable at 0 K. Based on the harmonic approximation, it has been suggested that CaH$_6$ becomes thermodynamically stable above approximately 1800 K; however, the transition to a liquid state was not explicitly considered. In our MLP-

MD simulations, both $CaH_{5.75}$ and $CaH_6$ transform into liquid-like states at approximately 1500–2000 K. As both systems approach similar liquid-like states at high temperatures, the free-energy difference between them is expected to diminish in the high-temperature limit.

Experimentally, $CaH_{5.75}$ has been reported to be synthesized at relatively low temperature[40], consistent with our MLP-MD results where $CaH_{5.75}$ emerges at around 1500 K. In the previous $CaH_6$ study[5], both $CaH_6$ and $CaH_{5.75}$ were observed after the synthesis under laser heating at around 2000 K. Although the interpretation differs from previous work[40], our results suggest that the reported formation of $CaH_6$ cannot be explained solely by thermodynamic stabilization; instead, kinetic factors—particularly Ca diffusion and precursor lattice compatibility—play a central role in selecting the reaction pathway. The precursor-dependent synthesis pathway presented here is consistent with all currently available experimental results, considering the thermodynamic stability of $CaH_{5.75}$ and $CaH_6$.

## V. CONCLUSIONS

In this study, we investigated the hydrogenation reactions leading to $CaH_6$ using machine-learning-potential molecular dynamics (MLP-MD) simulations. MD simulations at $CaH_x/H_2$ interfaces under 150 GPa revealed two competing hydrogenation pathways depending on the precursor phase. One pathway corresponds to the formation of A15-type $CaH_{5.75}$ at the $CaH_4/H_2$ interface. While this reaction is thermodynamically plausible, it requires substantial rearrangement of Ca atoms and therefore proceeds only at relatively high temperatures. The other pathway involves the formation of clathrate-type $CaH_6$ at the $CaH_2/H_2$ interface. Although $CaH_6$ is metastable, its Ca sublattice is closely related to those of other $CaH_x$ phases like $CaH_2$, enabling its formation even at lower temperatures where $CaH_{5.75}$ is kinetically inaccessible, through a diffusionless transformation reminiscent of a martensitic-like process.

A more quantitative discussion will require detailed experimental information on the local variation of temperature, sample morphology, and the location of the dominant reaction centers— for example, to determine whether hydrogenation occurs at bulk/$H_2$ interfaces or via hydrogen diffusion along polycrystalline grain boundaries. Reconciling these experimental conditions with the assumptions inherent in atomistic simulations will be essential for establishing practical synthesis guidelines. Nevertheless, the present results demonstrate that metastable superhydrides such as $CaH_6$ can be synthesized in practice by appropriately designing the precursor structural similarity, the H/M ratio, and temperature conditions. We expect that these insights will contribute to the rational design and experimental synthesis of new predicted superhydride materials.

More generally, this work demonstrates that machine-learning potential molecular dynamics simulations can be used to directly investigate competing reaction pathways and kinetic selection processes in complex materials systems under extreme conditions. This opens a new avenue for the use of machine-learning potentials, extending their applications beyond structure prediction to the exploration of synthesis pathways for realizing new materials.

## ACKNOWLEDGMENTS

This work was supported by JST GteX Program Japan Grant Number JPMJGX23H1, JST ASPIRE Japan Grant Number JPMJAP2421 and JSPS KAKENHI Grant-in-Aid for Early-Career Scientists, No. JP23K13542. The computation in this work was done using the facilities of the Supercomputer Center, the Institute for Solid State Physics, the University of Tokyo and supercomputing resources at Research Institute for Information Technology, Kyushu University.

*Ryuhei Sato: r-sato@g.ecc.u-tokyo.ac.jp

†Peter I. C. Cooke: pc729@cam.ac.uk

\* Ryuhei Sato: r-sato@g.ecc.u-tokyo.ac.jp

† Peter I. C. Cooke: pc729@cam.ac.uk

[40] W. Zhao, Q. Li, Y. Sun, Z. Wang, H. Li, H. Liu, H. Wang, Y. Xie, Y. Ma, arXiv:2512.03721

*Ryuhei Sato: r-sato@g.ecc.u-tokyo.ac.jp
†Peter I. C. Cooke: pc729@cam.ac.uk

# Competing Hydrogenation Pathways to Metastable CaH$_6$ Revealed by Machine-Learning-Potential Molecular Dynamics

*Ryuhei Sato,[1,*] Peter I. C. Cooke,[2,†], Maélie Caussé,[2,3] Hung Ba Tran,[3] Seong Hoon Jang,[3] Di Zhang,[3] Hao Li,[3] Shin-ichi Orimo,[3,4] Yasushi Shibuta[1], Chris J. Pickard[2,3]*

[1]Department of Materials Engineering, The University of Tokyo, Bunkyo-ku, Tokyo, Japan.
[2]Department of Materials Science and Metallurgy, University of Cambridge, Cambridge, United Kingdom
[3]Advanced Institute for Materials Research, Tohoku University, Sendai, Miyagi, Japan
[4]Institute for Materials Research, Tohoku University, Sendai, Miyagi, Japan

*Corresponding author. Email: rsato@material.t.u-tokyo.ac.jp (R. S.), pc729@cam.ac.uk (P. C.)

**Section S1**

To analyze the high temperature stability of $CaH_{5.75}$ and $CaH_6$ bulk phase, MLP-MD simulations of solid/liquid interfaces were analyzed at various temperatures. Here, liquid (or decomposed) $CaH_x$ phases were obtained by 50-ps MLP-MD simulations at 3000 K and 150 GPa. Then, the obtained "liquid" phases were contacted with bulk one and annealed for 1 ns at the specific temperatures. FIG. S7 shows the snapshots of the obtained structure after 1 ns. As shown in the snapshots, $CaH_6$ bulk were no longer maintained above 1500 K. On the other hand, $CaH_{5.75}$ bulk were observed below 1900 K, showing that this phase is relatively stable against high temperature. However, both of them were melted or decomposed at around 2000 K. These results suggest that, during laser heating, a molten or near-molten Ca–H state is likely to form in the vicinity of the laser-heated center.

**Table S1.** List of DFT-MD simulations used to construct MLP and validate the accuracy of MLP. MAE of H32 system was relatively large due to the *k* point setting difference.

| System (number of atoms) | T(K) | P(GPa) | time(1step=0.5fs) | MAE(MLP vs DFT)(eV/A) |
|---|---|---|---|---|
| Ca (Ca 8) | 1000, 3000 | 0 | 2.5 ps, 2.5 ps | 0.03 |
| $CaH_2$ (Ca 16 H 32) | 1000, 3000 | 0 | 3.4 ps, 2.0 ps | 0.07 |
| $CaH_2$ (Ca 16 H 32) | 10000 | 0 | 0.7 ps | N/A |
| $CaH_4$ (Ca 16 H 64) | 1000, 3000 | 25 | 3.2 ps, 2.8 ps | 0.13 |
| $CaH_4$ (Ca 16 H 64) | 10000 | 25 | 1.3 ps | N/A |
| $CaH_6$ (Ca 16 H96) | 1000, 3000 | 150 | 2.4 ps, 2.2 ps | 0.19 |
| $CaH_6$ (Ca 16 H96) | 10000 | 150 | 1.3 ps | N/A |
| $CaH_{24}$ (Ca 8 H192) | 1000, 3000 | 150 | 0.7 ps, 0.6 ps | 0.19 |
| $H_2$ (H 32) ※ *k* point = 8*8*4 | 1000 | 0 | 0.8 | 0.39 |
| $H_2$ (H 96) | 1000, 2000(only at 50 GPa) | 25, 50 | 2.5 ps, 2.5ps, 2.5 ps | 0.11 |
| $H_2$ (H 96) | 300, 600, 800, 1000, 1500, 3000 | 150 | 2.5 ps, 2.5 ps, 2.5 ps, 2.5 ps, 2.5 ps, 2.5 ps | 0.18 |
| $H_2$ (H 200) | 300, 1000, 3000 | 600 | 2.5 ps, 2.5 ps, 2.5ps | 0.16 |
| DFT-MD of $CaH_x/H_2$ interface | | | | |
| $CaH_2(100)/H_2$ interface (Ca16H160) | 1000, 3000 | 50 | 3.3 ps, 2.7 ps | 0.19 |
| $CaH_2(100)/H_2$ interface (Ca16H160) | 10000 | 50 | 2.1 ps | N/A |
| $CaH_4(001)/H_2$ interface (Ca 16 H 192) | 1000, 3000 | 50 | 3.5 ps, 2.4 ps | 0.19 |
| $CaH_4(001)/H_2$ interface (Ca 16 H 192) | 10000 | 50 | 2.3 ps | N/A |
| $CaH_6(100)/H_2$ interface (Ca16H288) | 1000, 3000 | 150 | 2.0 ps, 1.8 ps | 0.22 |
| **Not included in training** | | | | |
| $CaH_{5.75}$ (Ca64 H368) | 1000, 3000 | 150 | 1.0 ps, 0.9 ps | 0.20 |
| $CaH_{12}$ (Ca16 H192) | 1000, 3000 | 200 | 0.9 ps, 0.8 ps | 0.23 |

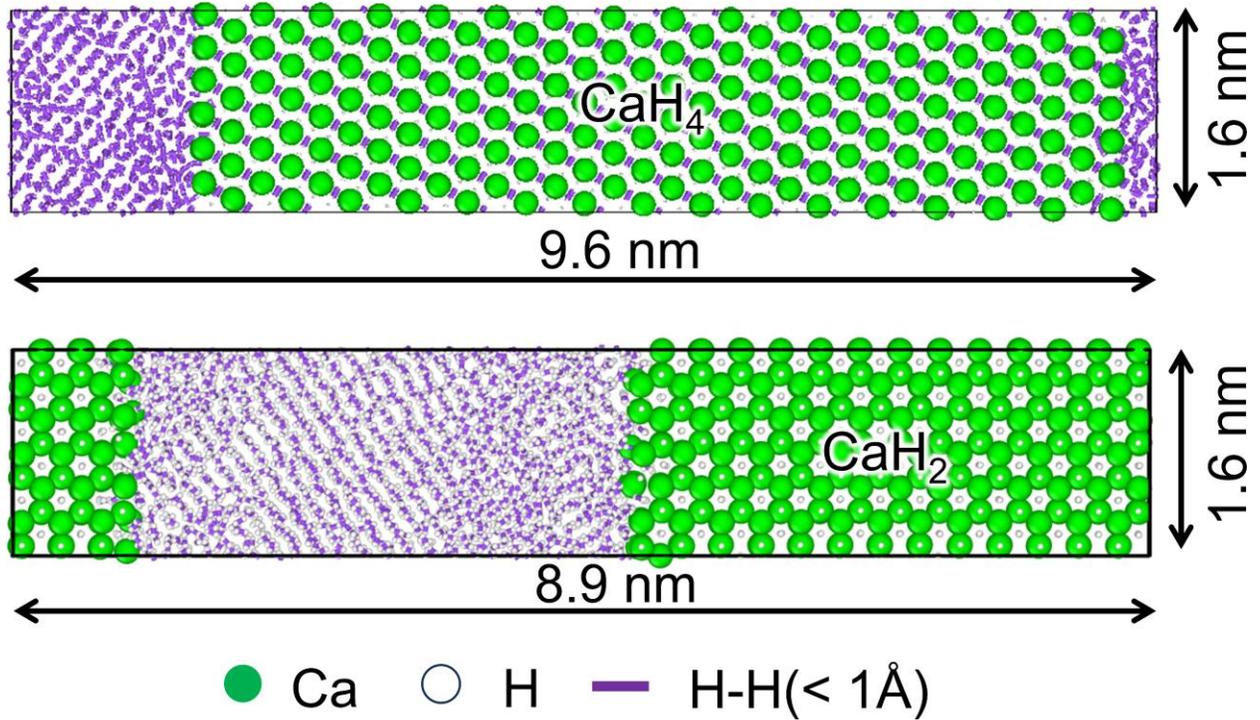

FIG. S1 Snapshots of the initial atomic configurations of $CaH_x/H_2$ interface used in these MLP-MD simulations

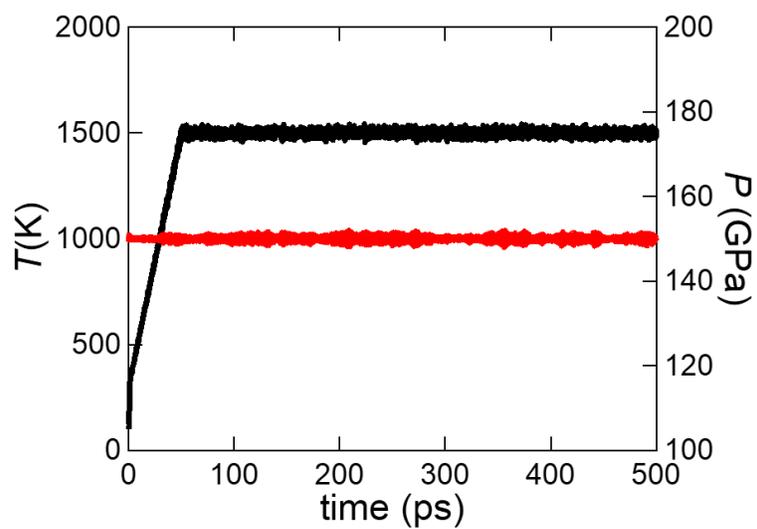

**FIG. S2.** Time series of Temperature and Pressure during MLP-MD simulation for CaH$_4$/H$_2$ interface. Note that the constant temperature and pressure (e.g. 15000 K and 150 GPa) were maintained during 1-ns MLP-MD simulation.

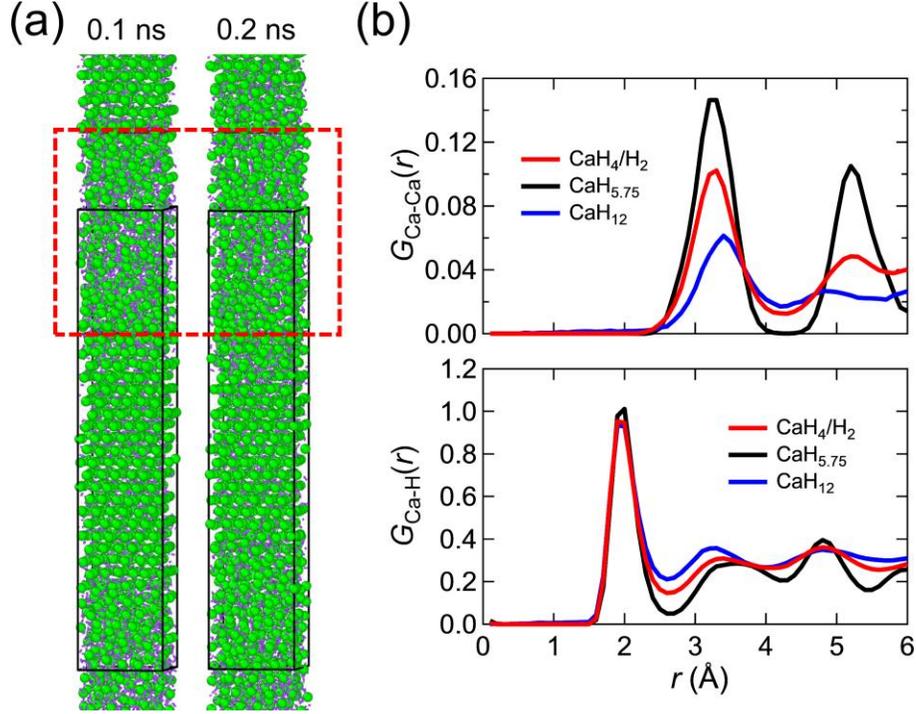

**FIG. S3.** (a) Snapshots of atomic configurations during the MLP-MD simulation of the $CaH_4(101)/H_2$ interface at 1500 K and 150 GPa. The red dashed line indicates the region in which the radial distribution function (RDF) was calculated to analyze the structure of the Ca-dissolved $H_2$ layer. (b) Time-averaged RDFs of the Ca-dissolved $H_2$ layer at the $CaH_4(101)/H_2$ interface (red solid line), as well as those of $CaH_{5.75}$ (black) and $CaH_{12}$ (blue) obtained from MLP-MD simulations at 1500 K and 150 GPa. Note that atomic configurations from 100 to 200 ps were used to compute the RDF of the Ca-dissolved $H_2$ layer.

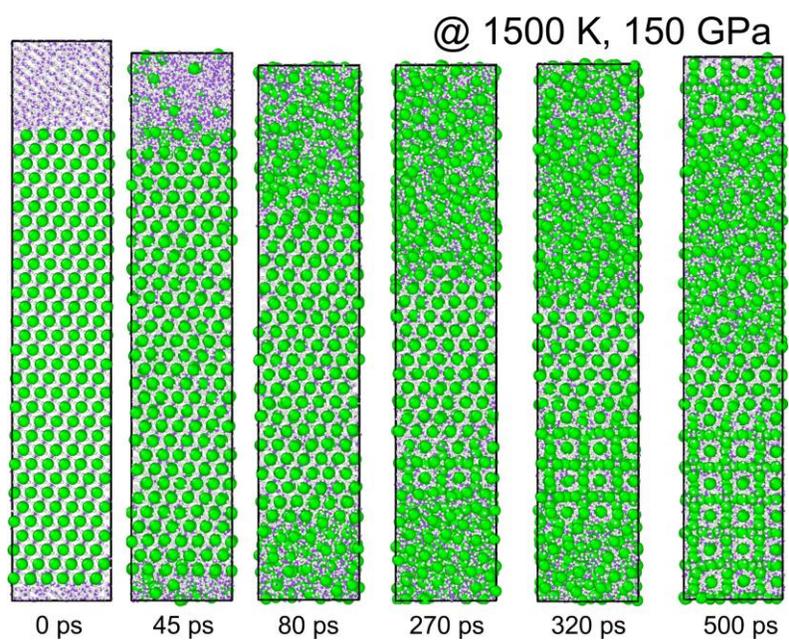

**FIG S4.** Snapshots of the atomic configurations of CaH$_4$(101)/H$_2$ interface during MLP-MD simulation at 1500 K and 150 GPa.

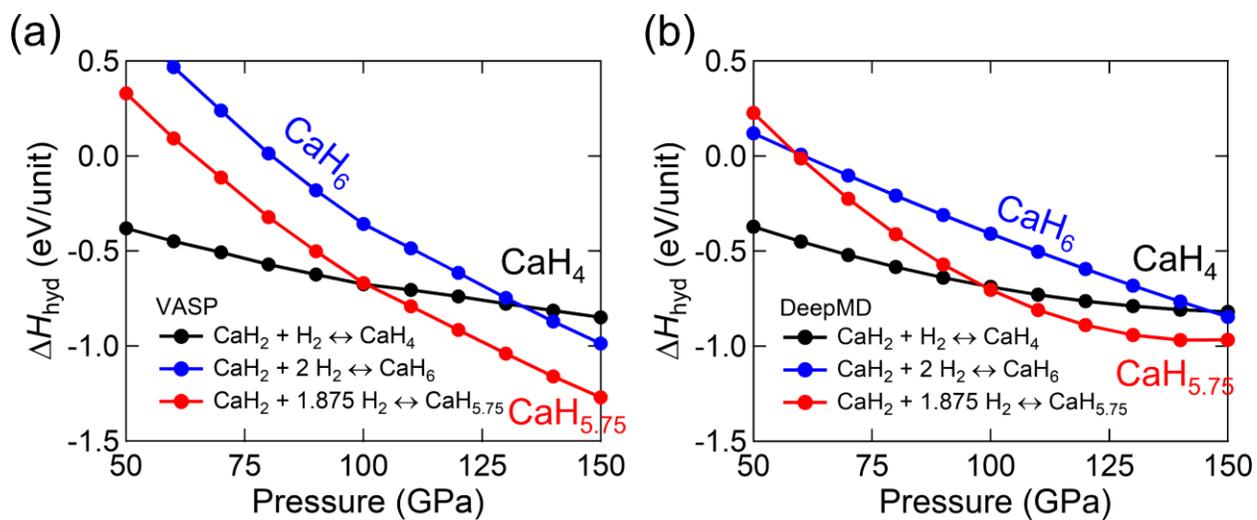

**FIG S5.** Hydrogenation enthalpy of $CaH_4$, $CaH_{5.75}$, and $CaH_6$ obtained from (a) DFT and (b) MLP calculation as a function of pressure.

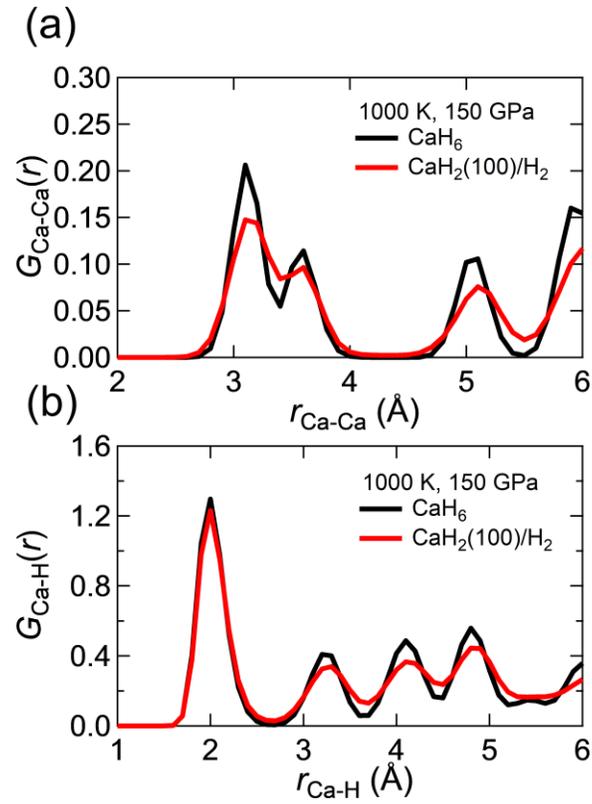

**FIG S6.** Time-averaged RDFs of the $CaH_x$ bulk phase inside the slab during MLP-MD simulation of $CaH_2(100)/H_2$ interface at 1000 K and 150 GPa, as well as those of $CaH_6$ obtained from MLP-MD simulations at 1000 K and 150 GPa. Note that atomic configurations from 0.9 to 1.0 ns were used to compute the RDF of the $CaH_x$ phase inside the slab.

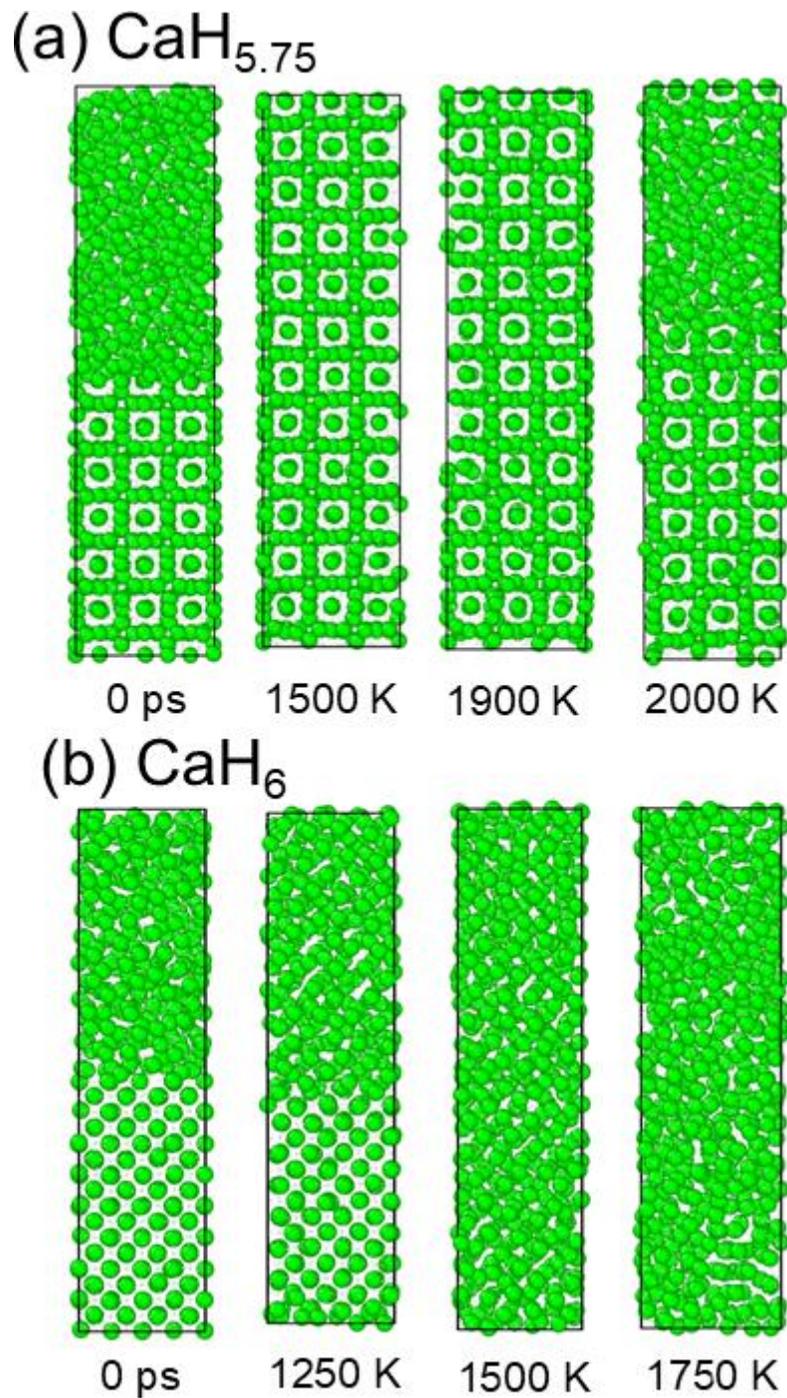

**FIG S7.** The snapshots of the atomic configurations during 1-ns MLP-MD simulations of (a) $CaH_{5.75}$ and (b) $CaH_6$ solid/liquid interface. The left most figures show the initial configurations, while others correspond to that after MLP-MD simulations at the corresponding temperatures. H-H bonds are not shown for the clarity.

**Movie S1.**

Movie for $CaH_{5.75}$ superhydride synthesis from $CaH_4(101)/H_2$ interface during 0.5-ns MLP-MD simulation at 1500 $K$ and 150 GPa.

**Movie S2.**

Movie for $CaH_6$ superhydride synthesis from $CaH_2(100)/H_2$ interface during MLP-MD simulation at 1000 $K$ and 150 GPa.